\documentclass[twocolumn,showpacs,preprintnumbers,amsmath,amssymb]{revtex4}

\usepackage{graphicx}
\usepackage{bm}

\begin{document}

\title{Orthorhombic phase of La$_{0.5}$Bi$_{0.5}$NiO$_{3}$ studied by first principles}

\author{J. Kaczkowski\footnote{corresponding author e-mail:kaczkowski@ifmpan.poznan.pl}, M. Pugaczowa-Michalska, A. Jezierski}
\affiliation{%
Institute of Molecular Physics, Polish Academy of Sciences\\M. Smoluchowskiego 17,
60-179 Pozna\'{n}, Poland
}%
\date{\today}

\begin{abstract}
The aim of presented first principles study of La$_{0.5}$Bi$_{0.5}$NiO$_{3}$ is to investigate electronic structure of orthorhombic phase \textit{Pbnm}. The calculations show that metallicity and magnetism of the system are strongly related with hybridization between Ni \textit{3d} and O \textit{2p}. To improve the quality of the electronic structure description of the system, especially the treatment of correlation for the Ni \textit{3d}, we employ GGA, LDA, and GGA+\textit{U}, LDA+\textit{U}. The LSDA results give good agreement with experiment. Thus, the screening effects originating from the hybridized \textit{3d} and O \textit{2p} electrons are sufficiently strong that they reduce the electronic correlations in the La$_{0.5}$Bi$_{0.5}$NiO$_{3}$, making it a weakly correlated metal.
\end{abstract}

\maketitle

\section{Introduction}
The bismuth perovskite oxides have attracted attention due to the different interesting properties like e.g. multiferroicity (BiFeO$_{3}$ \cite{Catalan, Chybczynska}), charge ordering (BiNiO$_{3}$ \cite{Azuma1,Azuma2}), or as a lead-free ferro- and piezoelectric materials (BiAlO$_{3}$ \cite{Kaczkowski}) to name the few. Variation of valence states can lead to notable phenomena such as superconductivity, magnetoresistance, negative thermal expansion in transition metal oxides. Much efforts has been focused on BiNiO$_{3}$ proved to undergo a phase transition accompanied by the charge redistribution between Bi and Ni ions \cite{Azuma1, Azuma2, Mizumaki, McLeod}. In ambient conditions, BiNiO$_{3}$ in triclinic structure (\textit{P-1}) displays the antiferromagnetic insulating ground state (\textit{G}-type AFM, T$_{N}$=300~K) \cite{Carl}, as well as exhibits an unusual charge distribution of Bi$^{3+}_{0.5}$Bi$^{5+}_{0.5}$Ni$^{2+}$O$_{3}$ \cite{Ishiwata1}. This charge disproportionation (CD) can be suppressed by applying an external pressure of several GPa or substituting Bi sites partially with La, which turns the system into the orthorhombic (\textit{Pbnm}) metallic phase \cite{Ishiwata2}. According to the analyses of X-ray absorption spectra for the orthorhombic phase of Bi$_{1-x}$La$_{x}$NiO$_{3}$ \cite{Wadati},  the valence of Ni ions is neither $+2$ nor $+3$ but in between of them, reflecting the nature of charge fluctuation between the A-site (i.e. Bi) and the B-site (i.e. Ni) ions. The end members of the solid solution (i.e. BiNiO$_{3}$ and LaNiO$_{3}$) have differing ground state electronic properties. LaNiO$_{3}$ is a metal and the only member of the perovskite nickelates family (RNiO$_{3}$, where R is a rare earth) lacking any magnetic order in its bulk form.

Our previous studies \cite{MPM} revealed the complex nature of the insulating band gap in BiNiO$_{3}$ in the \textit{P-1} structure that arises not only from the correlation effect of Ni \textit{3d} orbitals, but also from CD at Bi sites. The correlation effect of Ni \textit{3d} was described by DFT+$U$ method. This method is generally regarded to be the most computationally feasible means to reproduce the correct insulating ground states in correlated systems. The theoretical description of reduction in CD at Bi sites by La substitution at Bi (i.e. Bi$^{5+}$ site) of BiNiO$_{3}$  supports the decrease in the band gap value. However, that it is not sufficient to close the insulator band gap in the system in \textit{P-1} structure.
Here we present the results of orthorhombic phase of  La$_{0.5}$Bi$_{0.5}$NiO$_{3}$ studied by first principles. The orthorhombic \textit{Pbnm} structure is the ground state of La$_{0.5}$Bi$_{0.5}$NiO$_{3}$ \cite{Ishiwata3}. This issue will be discussed along with the connection of our results with the experiments \cite{Ishiwata3,Wadati}.
\section{Method of calculations}
The calculations are performed using the projector augmented wave method (PAW) \cite{Blochl, Kresse1}, which is implemented in the Vienna ab initio Simulation Package (VASP) \cite{Kresse2}. The Perdew-Burke-Ernzerhof (PBE) \cite{Perdew1} generalized gradient and Perdew-Zunger (PZ) \cite{Perdew2} local density approximations (LDA) with and without on-site Coulomb interactions (DFT+\textit{U}) were used for the exchange-correlation potentials. The DFT+\textit{U} method is generally employed to reproduce the correct ground states in correlated systems (like an insulator or semiconductor). The typical application of the DFT+\textit{U} method introduces a correction to the DFT energy by introducing a single numerical parameter \textit{U}. The value of the on-site Coulomb repulsion parameter was chosen as \textit{U} = 7 eV for Ni \textit{3d} (as for BiNiO$_{3}$). In the frame of the DFT+\textit{U} we have used the rotationally invariant approach of \cite{Dudarev}.
\section{Results}
We performed the total energy calculations of Bi$_{0.5}$La$_{0.5}$NiO$_{3}$ for various possible occupations of Bi and La sites in two crystallographic phases: triclinic \textit{P-1} and orthorombic \textit{Pbnm}. Our calculations show that the total energy of Bi$_{0.5}$La$_{0.5}$NiO$_{3}$ in the orthorhombic \textit{Pbnm} structure is lower that the total energy of the system in the triclinic \textit{P-1} phase for both DFT and DFT+\textit{U} approaches. The differences between the total energies of these structures are at least $41.8~meV/f.u.$ and $36~eV/f.u.$  from DFT and DFT+\textit{U} approaches, respectively. 

\begin{figure}
\includegraphics[scale=0.06]{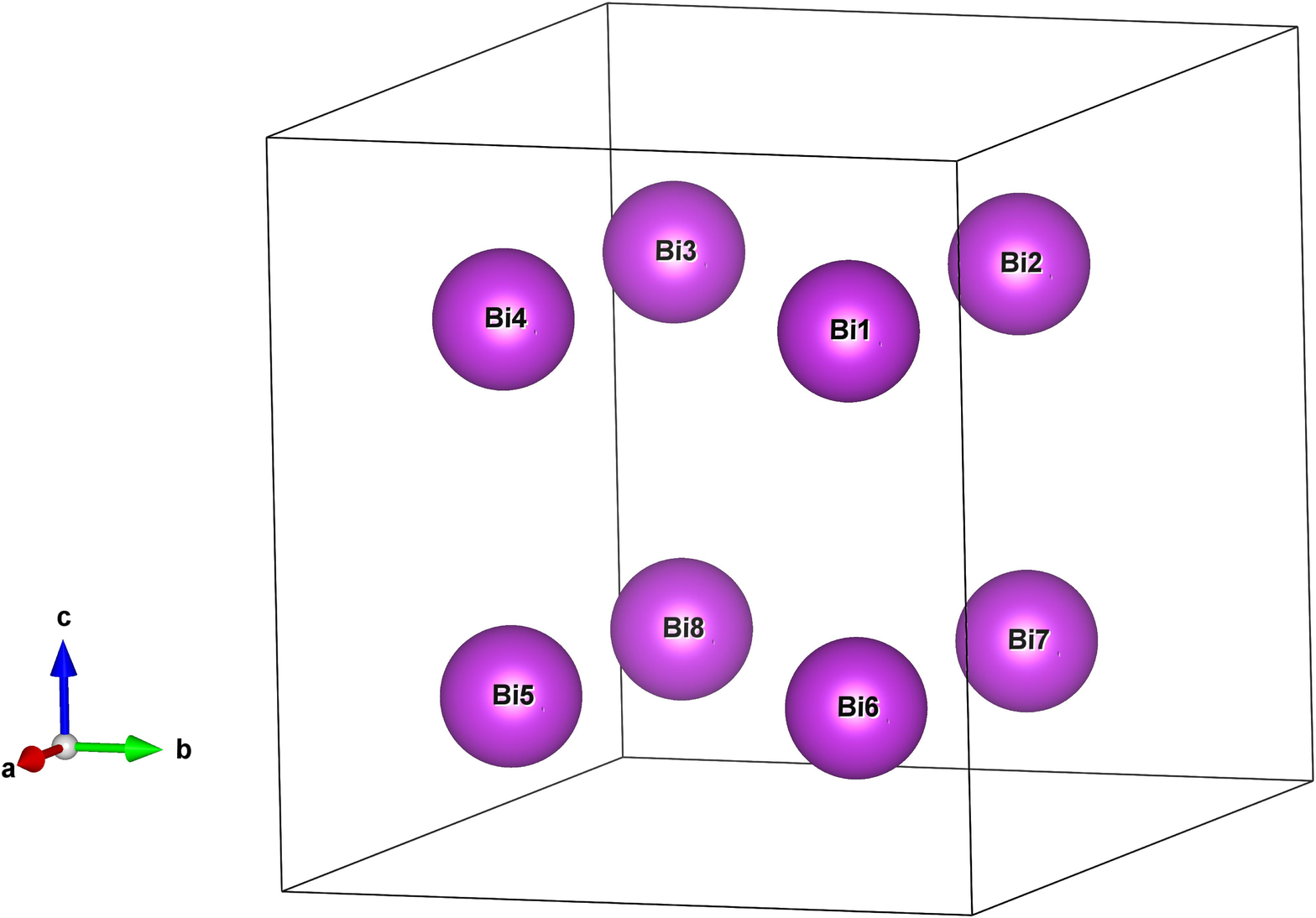}
\caption{\label{fig:epsart-1} A sketch of Bi sublattice in our study of Bi$_{0.5}$La$_{0.5}$NiO$_{3}$ whithin the \textit{Pbnm} phase. Eight position are denoted as Bi1, Bi2, Bi3, Bi4, Bi5, Bi6, Bi7 and Bi8.}
\end{figure}

In order to study the site preference of La ion in BiNiO$_{3}$ in the low-energy \textit{Pbnm} structure we substitute some of Bi atoms in the supercell of the parent system by La. In Bi$_{0.5}$La$_{0.5}$NiO$_{3}$ the effect of the substitutional disorder was studied by simulating explicitly all inequivalent Bi/La arrangements within the 40-atom cell of orthorombic phase. Fig.\ref{fig:epsart-1} shows a sketch of Bi sublattice in our study of Bi$_{0.5}$La$_{0.5}$NiO$_{3}$. Six inequivalent Bi/La arrangements are possible for eight positions in the Fig.\ref{fig:epsart-1}. According to the sketch, La ions may occupy the following Bi positions: (1,3,6,8), (2,4,6,8), (4,5,6,8), (1,2,3,4), (3,5,6,8) or (2,5,6,8). Moreover, four types of magnetic ordering at Ni cations sites are considered for the studied orthorombic phase: an ferromagnetic (\textit{FM}) one and three antiferromagnetic (\textit{A-}, \textit{C-}, and \textit{G-AFM}) ones. The detailed description of these spin arrangements of Ni sublattices are the same as in \cite{MPM}. For all the Bi$_{0.5}$La$_{0.5}$NiO$_{3}$ confgurations investigated, the atomic structure was fully relaxed until residual forces become smaller than $0.01~eV/\AA$. As a general rule, we note that the system prefers Bi/La arrangements that respect the "rocksalt order", i.e. (2,4,6,8). Furthermore, all of the employed exchange-correlation approaches do not change this feature. Others Bi/La arrangement of Bi$_{0.5}$La$_{0.5}$NiO$_{3}$ correspond to high energy ($>22~meV/f.u.$ for DFT and $>19~meV/f.u.$ for DFT+\textit{U}) and therefore, we focused on the Bi/La arrangement of Bi$_{0.5}$La$_{0.5}$NiO$_{3}$ with the "rocksalt order". 

Experimental investigations of Bi$_{1-x}$La$_{x}$NiO$_{3}$ have shown that the sample with $x=0.5$ is metallic down to 5 K \cite{Ishiwata3}. Temperature dependence of the inverse magnetic susceptibility for the same sample with $x=0.5$ is more consistent with $S=1/2$ (Ni$^{3+}$) \cite{Ishiwata3}. We found that LDA functional reproduces the experimental metallic ground-state (GS) with $0~\mu_{B}$ on Ni in Bi$_{0.5}$La$_{0.5}$NiO$_{3}$. For calculations with PBE functional we were enabled to achieve metalic GS of the system for low energy \textit{FM} ordering, but the magnetic moments on Ni are in range of $0.378 \div 0.582~\mu_{B}$. The consequence of including correlation on \textit{3d} shell of Ni atom for both LDA+\textit{U} and PBE+\textit{U} calculations is that the GS of the system in \textit{Pbnm} is the insulating with the band gap at least of $0.403~eV$. The \textit{G-AFM} order is preferable in DFT+\textit{U} due to the value of total energy of the Bi$_{0.5}$La$_{0.5}$NiO$_{3}$. We note, that this configuration, however, has not been experimentally reported for Bi$_{1-x}$La$_{x}$NiO$_{3}$ with $x \ge 0.2$ in \cite{Ishiwata3,Wadati}.
\begin{figure}
\includegraphics[scale=0.5]{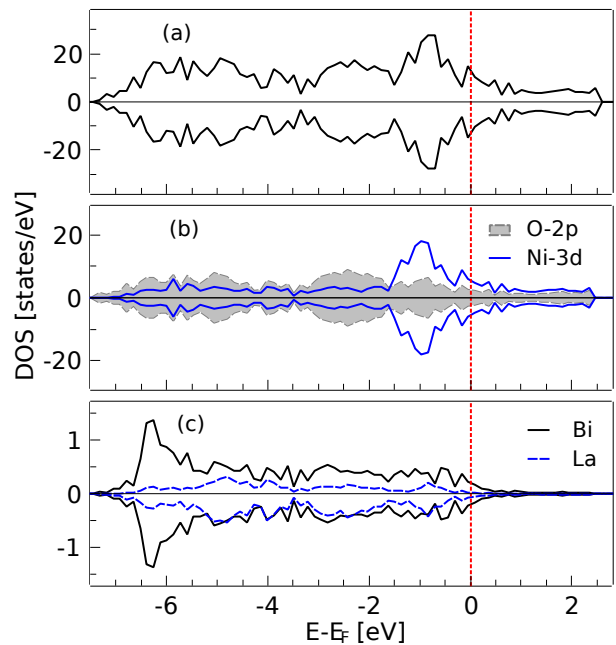}
\caption{\label{fig:epsart-2} The density of states (DOS) of Bi$_{0.5}$La$_{0.5}$NiO$_{3}$ in \textit{Pbnm} obtained from LDA: (a) the total DOS; (b) the DOS of Ni-\textit{3d} and O-\textit{2p}; (c) the DOS of Bi and La ions. The vertical line is Fermi energy.}
\end{figure}

\begin{figure}
 \includegraphics[width=0.5 \textwidth]{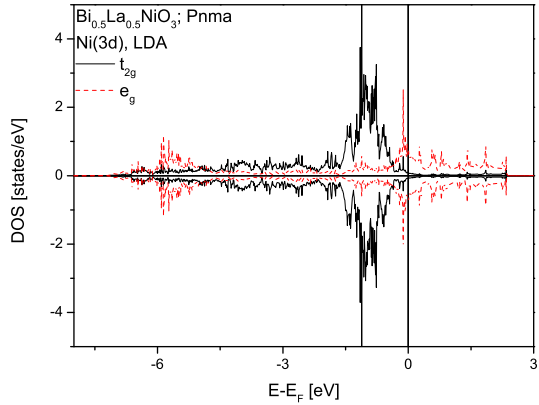}
 \caption{\label{fig:epsart-3}The DOS of $e_{g}$ and $t_{2g}$ states of Ni.}
\end{figure}

The charge disproportionation (CD) of Bi ions into Bi$^{3+}$ and Bi$^{5+}$ in the triclinic phase of BiNiO$_{3}$ has been discussed by the electron localization function in \cite{MPM}. Our previous studies on BiNiO$_{3}$ revealed that the CD of Bi is suppresed by La ions within \textit{P-1} structure. However, total energy calculations presented here confirmed experimental results of Ishiwata et al. \cite{Ishiwata1,Ishiwata2,Ishiwata3} that random substitution of La for Bi causes transition to higher symmetry, i.e. \textit{Pbnm}. The experimental results were interpreted as a transfer of the oxygen holes from the Bi---O sublattice to the Ni---O sublattice to be expressed as Bi$^{3+}+$Bi$^{5+}+$2Ni$^{2+}$ (triclinic) $\rightarrow$ 2Bi$^{3+}+$2Ni$^{2+}\underline{L}$ (orthorhombic) \cite{Wadati}. Thus, we focus on results of electronic structure calculations, especially on orbital occupancy of electrons. The main motive of the crystal \textit{Pbnm} structure are NiO$_{6}$ octahedra which are connected via all vertices. As a result the bands are broader and the electron states more itinerant.  Fig.\ref{fig:epsart-2} presents the density of states (DOS) of Bi$_{0.5}$La$_{0.5}$NiO$_{3}$ in \textit{Pbnm} phase calculated using LDA approach. The largest part of the valence band within the energy range $-7~eV$ to $0~eV$ is dominated by O-\textit{2p} and Ni-\textit{3d} states with some small contributions from Bi and La electronic states. From the partial DOS of O-\textit{2p} and Ni-\textit{3d} (Fig.\ref{fig:epsart-2}(b)), one can see that the Fermi level comes through the O-\textit{2p} states having significant admixture of Ni-\textit{3d}, especially $e_{g}$-like orbitals (Fig.\ref{fig:epsart-3}). In the studied system the localized Ni-\textit{3d} states (mainly $t_{2g}$-like) centered at 1.2 eV below the Fermi level are visible as a strong peaks in atom-resolved DOS for spin-$\uparrow$ and spin-$\downarrow$ directions. A set of delocalized Ni \textit{3d} states (mostly $e_{g}$-like) crosses the Fermi level. The Ni $t_{2g}$ orbitals are occupied for both spin-$\uparrow$ and spin-$\downarrow$ DOS and found in energy range $-6.5$ to $0.0~eV$. The Ni $e_{g}$ states are partially filled. A sharp peak is visible close to the Fermi level. From inspection of projected DOS obtained by LDA we may infer that in the Bi$_{0.5}$La$_{0.5}$NiO$_{3}$ (with \textit{Pbnm}) the valence of Ni behaves approximately as $(2+\delta)+$, rather than $3+$.
 
The computed structural parameters of Bi$_{0.5}$La$_{0.5}$NiO$_{3}$ in \textit{Pbnm} structure are given in Table 1.
\begin{table}
\caption{Calculated lattice constant \textit{a}, orthorhombic angles, volume of Bi$_{0.5}$La$_{0.5}$NiO$_{3}$ in \textit{Pbnm} phase}
\begin{tabular}{ c l l l l l l} \hline
 LDA: & $a=5.3214$ \AA, $b=5.4022$ \AA, $c=7.4028$ \AA &\\
  & $\alpha  = \beta = 90.0^{0}$, $\gamma = 89.1365^{0}$, $V=212.815$ \AA$^{3}$&\\
\hline
\end{tabular}
\end{table}

\section{Conclusions}
Our study has shown that the presence of $50\%$ La in BiNiO$_{3}$, as well as random substitution of La by Bi leads to higher symmetry of the system. The calculated total energy of Bi$_{0.5}$La$_{0.5}$NiO$_{3}$ in the orthorombic \textit{Pbnm} structure is lower than one in the triclinic \textit{P-1}. This result is in agreement with experimental evidence. The metalic ground state is obtained for LDA and PBE calculations. The Coulomb repulsion in Ni-\textit{3d} orbital for both LDA+\textit{U} and PBE+\textit{U} calculations ($U=7~eV$) provides the insulating ground state in Bi$_{0.5}$La$_{0.5}$NiO$_{3}$ with \textit{G-AFM} magnetic ordering. Based on our calculations, we show that the electronic screening effect from the delocalized Ni-\textit{3d} and O-\textit{2p} states should mitigate the electronic correlations of Ni atoms, making Bi$_{0.5}$La$_{0.5}$NiO$_{3}$ a weakly correlated metal.

\def\refname{References}

\end{document}